\documentclass[prl,preprint,endfloats]{revtex4-1}
\usepackage{bm,graphicx,graphics,amsmath,amssymb,bm,epsfig,color}
\usepackage{euscript,tabularx}
\usepackage{longtable}

\begin{document}

\title{Conduction of spin currents through insulating oxides}

\author{Christian Hahn} 
\affiliation{Service de Physique de l'\'Etat Condens\'e (CNRS URA
  2464), CEA Saclay, 91191 Gif-sur-Yvette, France}

\author{Gr\'egoire de Loubens} 
\affiliation{Service de Physique de l'\'Etat Condens\'e (CNRS URA
  2464), CEA Saclay, 91191 Gif-sur-Yvette, France}

\author{Vladimir V. Naletov} 
\affiliation{Service de Physique de l'\'Etat Condens\'e (CNRS URA
  2464), CEA Saclay, 91191 Gif-sur-Yvette, France}
\affiliation{Institute of Physics, Kazan Federal University, Kazan
  420008, Russian Federation}

\author{Jamal Ben Youssef}
\affiliation{Universit\'e de Bretagne Occidentale, Laboratoire de
  Magn\'etisme de Bretagne CNRS, 6 Avenue Le Gorgeu, 29285 Brest,
  France}

\author{Olivier Klein} 
\affiliation{Service de Physique de l'\'Etat Condens\'e (CNRS URA
  2464), CEA Saclay, 91191 Gif-sur-Yvette, France}

\author{Michel Viret} 
\affiliation{Service de Physique de l'\'Etat Condens\'e (CNRS URA
  2464), CEA Saclay, 91191 Gif-sur-Yvette, France}

\date{\today}

\begin{abstract}

  \textbf{Spintronics is a field of electronics based on using the
    electron spin instead of its charge. The recent advance in the
    manipulation of pure spin currents, i.e. angular momentum transfer
    not associated to conventional charge currents, has opened new
    opportunities to build spin based devices with low energy
    consumption \cite{jungwirth12}. It has also allowed to integrate
    ferromagnetic insulators in spintronic devices, either as spin
    sources \cite{kajiwara10,sandweg10,castel12a,hahn13,kelly13} or
    spin conductors \cite{chumak12,kajiwara10,xiao12} using their
    magnetic excitations to propagate a spin signal. Antiferromagnetic
    insulators belong to another class of materials that can also
    sustain magnetic excitations, even with a higher group velocity
    \cite{KittelBook}. Hence, they have potential as angular momentum
    conductors, possibly making faster spin devices. At the opposite
    end, angular momentum insulators are also required in spintronic
    circuits. The present letter underlines some essential features
    relevant for spin current conduction, based on measurements of
    angular momentum transmission in antiferromagnetic NiO and in the
    non-magnetic light element insulator SiO$_2$.}

\end{abstract}

\maketitle

Spin conductors are materials through which a spin current can flow.
These are conventionally normal metals where conduction electrons
provide the necessary mobile spinned entities. The added properties of
light mass and low resistivity have made Al one of the best spin
conductors so far \cite{valenzuela06,kimura07}, with spin diffusion
lengths as large as one micron.  Another alternative to propagate a
spin current is to use spin excitations in magnetically ordered
materials.  Magnons are the elementary excitations in ferromagnets
(FM), each reducing the total spin by one unit of $\hbar$, and hence
the magnetic moment by $g \mu_B$ \cite{KittelBook} where $\mu_B$ is
the Bohr magneton and $g$ the Land\'e factor.  They can propagate over
distances in the millimetre range in the insulating ferrimagnet
Yttrium Iron Garnet (YIG), which has unsurpassed low damping
\cite{kajiwara10,sandweg10,castel12a,
  hahn13,kelly13,chumak12}. Therefore, magnetic insulators represent
an attractive medium for propagating spin information over long
distances.  However, these spin excitations have a direction imposed
by that of the magnetization, which can also potentially couple to
that of other magnets in the circuit through direct contact or stray
fields. Non magnetic elements are not restricted by these limitations.
Interestingly, elementary magnetic excitations also exist in
antiferromagnets (AF), materials with two magnetic sublattices
oppositely oriented, resulting in a global zero moment
\cite{KittelBook}. Antiferromagnetic magnons, like their ferromagnetic
counterparts, correspond to the reversal of a single spin in an AF
crystal. Thermally generated AF magnons carry random spin directions
resulting in zero global angular momentum, but each of these entities
can transport spin information on significant distances and faster
than in ferromagnets.  We intend here to demonstrate that angular
momentum can be transmitted through an AF insulator inserted between a
spin source and a spin sensor.

Spin sources have conventionally relied on extracting spin polarized
conduction electrons from ferromagnetic metals. This can be achieved
by driving a charge current through the ferromagnet or applying
thermal gradients (Peltier effect) \cite{slachter10}. A non-local
geometry can also be designed so as to extract a pure spin current
'unpolluted' by charge currents \cite{jedema01,valenzuela06}.  A
contactless means of emitting angular momentum can also be achieved
using a FM undergoing ferromagnetic resonance (FMR). Damping processes
generate spin currents at the interface with a normal metal, via a
phenomenon known as spin pumping \cite{tserkovnyak05}. This recent
discovery has far reaching implications as magnetic insulators can now
be used as active elements in spintronic devices.  In particular,
Yttrium Iron Garnet is the material of choice
\cite{kajiwara10,sandweg10,castel12a, hahn13,kelly13,chumak12} and the
one we selected as our radio-frequency driven spin source.

At the opposite end of the circuit is the spin-sensor, relying on the
conversion of a spin current into a charge current, through a
mechanism known as the inverse spin Hall effect
(ISHE)\cite{valenzuela06,kimura07}. The effect is based on spin-orbit
coupling, as originally described for the direct spin Hall effect
\cite{dyakonov71,hirsch99}, and it enables spin current sensing,
including those emitted from a ferromagnet undergoing FMR
\cite{saitoh06,kajiwara10}. The best ISHE materials are heavy element
metals, for which spin-orbit effects are enhanced. Pt is the most
commonly used material, and the one we chose here, but others like Ta
or W are also promising candidates \cite{Liu12,hahn13,pai12,wang13}.

The device we design is thus a simple tri-layer with spin source and
sensor separated by the material through which spin conduction is to
be measured. We propose here to study nickel oxide, a well known
antiferromagnetic insulator with a N\'eel point well above
room-temperature \cite{Hutchings72}.  A series of tri-layers
YIG$|$NiO$|$Pt were grown on (111) GGG substrates, each with a single
crystalline 200 nm thick film of YIG (deposited by Liquid phase
Epitaxy) and a 5 nm thick Pt layer. The NiO layers have thicknesses
ranging from zero to 30 nm and were deposited under a 300 Oe magnetic
field. Importantly, magnetometry measurements evidence the absence of
exchange bias in our trilayers, whose coercivity is around 0.5 Oe,
totally unchanged from that of the bare YIG film. This demonstrates
that NiO is magnetically decoupled from the YIG, unlike what is
systematically found for the NiO$|$Ni system \cite{Jamal07}.  The
samples are then mounted on a 500~$\mu$m wide, 2~$\mu$m thick Au
transmission line cell used for microwave generation up to 20~GHz. The
long axis of the 5 mm $\times$ 1 mm samples is mounted parallel to the
excitation field $h_\text{rf}$ as indicated in the inset of Fig.1. The
inverse spin Hall voltage $V_\text{ISH}$ in Pt is measured by a
lock-in technique (with the microwave power turned on and off at a
frequency of a few kHz) with electrical connections through gold leads
on each side of the area of excitation. When in ferromagnetic
resonance, YIG emits a spin current at the interface with NiO, itself
in contact on the other side with the Pt layer converting spin into
charge.

The ISHE generated in the reference YIG$|$Pt sample is shown in
Fig. 1-a, evidencing the significant spin current emitted when the YIG
enters resonance, as already reported \cite{hahn13}. Measured Hall
voltages are odd in field and can reach here $27~\mu$V at 3.85 GHz for
an output power of 10 dBm (i.e. a radio-frequency field of 0.2
Oe). The effect can be understood using three key parameters: the spin
mixing interface conductance $G_{\uparrow \downarrow} \simeq
10^{14}~\Omega^{-1}\cdot$m$^{-2}$, the spin Hall angle
$\theta_{SH}\simeq 0.05$ and the spin diffusion length
$\lambda_{sd}^\text{Pt}=2$~nm. These are representative of the physics
at the heart of the effect. The dynamically generated spin current is
emitted through the interface which acts as a semi-transparent medium
described by the 'conductance' $G_{\uparrow \downarrow}$. The exact
nature of this parameter is at present not completely clear, but
experimentally, metallic interfaces have a much better transparency
than insulator/metallic ones \cite{Czeschka11}. This probably reflects
the importance of a common vector of spin transport, being the
conduction electrons in this specific case. The second important
quantity is $\lambda_{sd}$, the distance on which spin memory is
preserved in the normal metal, set by the by the spin-flip scattering
rate. It is rather short in Pt where spin currents are efficiently
converted into charge currents by the inverse spin Hall effect. The
figure of merit of this process is the spin Hall angle $\theta_{SH}$,
determined by the difference in average scattering angle between spin
up and spin down electrons.

Once the angular momentum source and detector tested using the
YIG$|$Pt structure, the effect of inserting the NiO layer can be
studied.  For NiO thicknesses ranging from 2 nm to 15 nm a ISHE
voltage can be measured in the Pt layer as shown in Fig. 2-a. We argue
that this is the conceptual demonstration that antiferromagnets can
indeed conduct angular momentum. The amplitude of the ISHE signal is
however much lower than in the reference YIG$|$Pt sample (compare
Fig. 1-a and 1-b), thus evidencing the low transparency of the
YIG$|$NiO interface and/or the imperfect transmission of NiO. It is
therefore interesting to study the evolution of the angular momentum
transfer as a function of NiO thickness, as shown in Fig. 2. The spin
current signal is found to decrease in an exponential fashion leading
to an estimated diffusion length around 2 nm. This is extremely small
compared to what one could expect if magnons were propagating through
the antiferromagnet. Instead, this corresponds more to the actual size
of these entities, hinting to a non propagative angular momentum
transport. We would like to point out here that this dependence in NiO
thickness rules out an explanation for the effect based on conduction
electron tunnelling \cite{HammelCondMat}, or purely on the magnetic
susceptibility of NiO \cite{KittelBook,Hutchings72}. Indeed, the NiO
susceptibility at GHz frequencies is low, it should not vary with
thickness and a hypothetical amplification at the YIG interface cannot
be expected considering the absence of magnetic coupling. Another very
interesting experimental observation in Fig. 2 is the decrease of spin
conductance for the thinnest 2 nm thick layer. We underline here that
this is a real effect as any problem of layer discontinuity would
instead significantly enhance the signal through the regions with
direct YIG$|$Pt contact. Reduced magnetic quality or/and spatial
confinement induced overlap of antiferromagnetic magnons might be
responsible for this loss of angular momentum transfer. Clearly, the
behaviour of these entities in these thin layers need to be
theoretically studied in the future.  The transparency of the
YIG$|$NiO interface can be inferred from the extrapolation of the
exponential behaviour in Fig. 2 to zero thickness, which is found to
be twenty times smaller than the ISHE voltage in the YIG$|$Pt
reference sample. Thus, the total $G_{\uparrow \downarrow} \simeq
5.10^{12}~\Omega^{-1}\cdot$m$^{-2}$, which can be interpreted as
resulting from the spin resistances of the two interfaces in
series. We argue that the critical interface is the YIG$|$NiO one
where the transfer of angular momentum cannot rely anymore on
exchanging conduction electrons. Instead, the magnons in YIG have to
transform into antiferromagnetic magnons in NiO, which are to be found
on average at a much higher frequency/energy and with a dispersion
linear in wave vector \cite{KittelBook,Hutchings72}. This mismatch is
likely to be the bottleneck of the transmission problem but this is
not demonstrated here and should be tested in the future. Moreover,
the growth of micro-crystalline NiO on top of the excellent quality
YIG crystal are very likely not optimal. Hence, the YIG$|$NiO
interface transparency to the diffusion of angular momentum can be
estimated to be around 0.05 (attributing the original $G_{\uparrow
  \downarrow}$ value to the NiO$|$Pt interface). There is an obvious
need for a deeper theoretical understanding of the problem, but this
is beyond the scope of the present study.

Beside the availability of good angular momentum conductors, the
design of useful devices also requires the identification of good
insulators. These should be non-magnetically ordered in order to
suppress the magnon channel. However, this is probably not enough as,
although often forgotten, phonons are also able to carry angular
momentum \cite{kurebayashi11} (like in the Einstein-de Haas effect),
especially in materials with strong spin-orbit coupling
elements. Therefore, light oxides, free from the existence of magnetic
moments and spin-orbit interactions, look like good candidates. Here
we select SiO$_2$. Remarkably, as shown in Fig. 3, SiO$_2$ layers as
thin as 2 nm are able to completely suppress the spin current
propagation in our YIG$|$SiO$_2|$Pt trilayers. This is an impressive
effect showing that angular momentum cannot even 'tunnel' through such
a thin layer. Thus, SiO$_2$ can be considered an excellent angular
momentum insulator, which we argue comes from the absence of magnetic
excitations in this compound, together with the light O and Si atoms
reducing spin-orbit interactions. Beyond these concepts, it would be
very interesting to study in more details the behaviour of phonons
through the interface and their role in conducting angular momentum.

In conclusion, we show here that antiferromagnets can be classified as
spin conductors, probably because their antiferromagnetic magnons can
propagate angular momentum. On the other hand, light non magnetic
insulators are good spin insulators because their phonons are
inefficient to carry orbital momentum. The diffusion length measured
in antiferromagnetic NiO is however inconsistent with a picture of
mobile magnons diffusing from the spin source to the spin
sensor. Instead, it seems that angular momentum crosses through
confined non-propagating antiferromagnetic magnons. The YIG$|$NiO
interface transparency is found to be of the order of 5 \%, possibly
reflecting the magnon energy mismatch as well as the non optimal
interface quality as NiO is not epitaxially grown on single
crystalline YIG. Therefore, beyond the conceptual advance brought by
the present results, there is room for optimization on the materials'
side and a critical need for theoretical understanding of angular
momentum transfer through interfaces in the absence of conduction
electrons.


\begin{figure}
 \includegraphics[scale=0.5, angle=0]{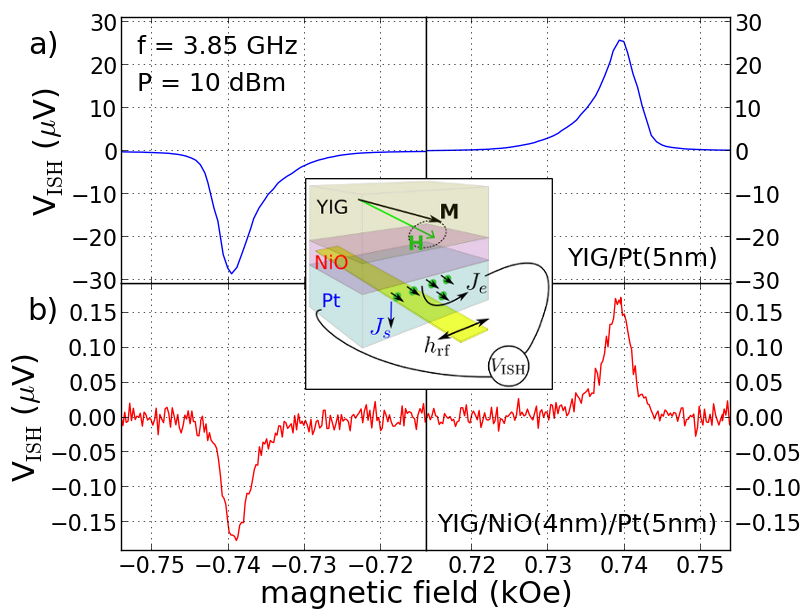}
 \caption{(a): ISHE signal measured in a YIG$|$Pt bilayer generated by
   the spin current emitted by the YIG at resonance at a frequency of
   3.85 GHz and a power of 10 dBm. (b) : Similar signal when a 4 nm
   NiO layer is inserted between YIG and Pt. Note the difference in
   scale for the measured ISHE voltage. Inset: geometry of the
   measurement showing the induced spin current in the tri-layer
   geometry.}
 \label{fig1}
\end{figure}

\begin{figure}
 \includegraphics[scale=0.5, angle=0]{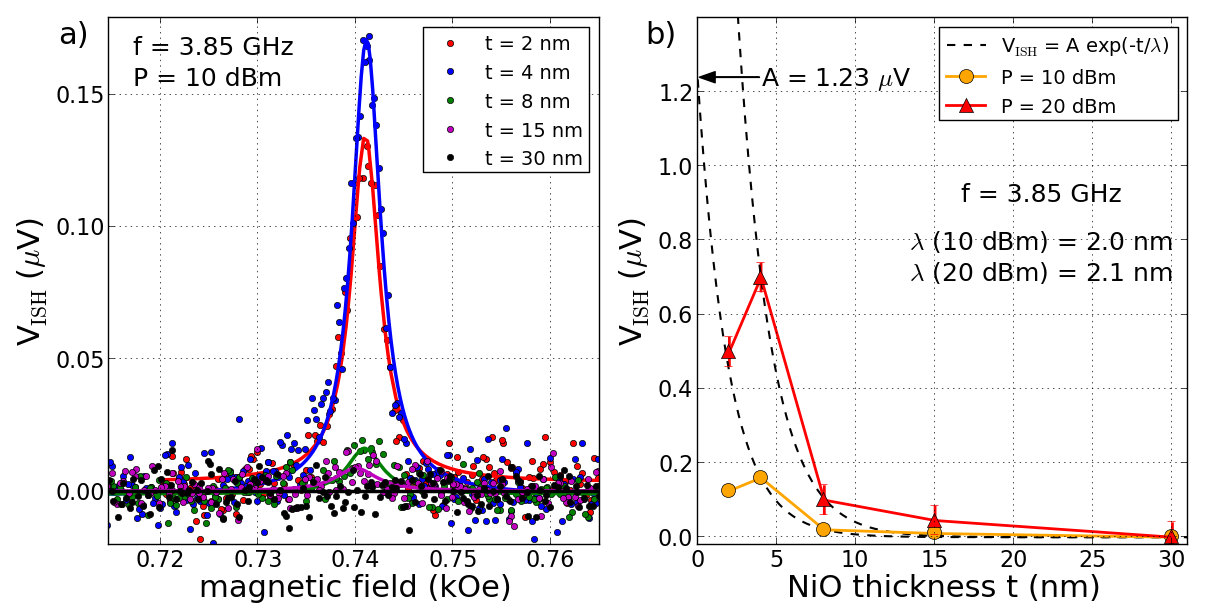}
 \caption{NiO thickness dependence of the angular momentum transfer in
   YIG$|$NiO(t)$|$Pt devices at 3.85 GHz and two different powers: 10
   and 20 dBm. (a): Raw measurements at 10 dBm and positive
   field. (b): Summary of the NiO thickness dependence of the measured
   spin Hall Voltage at 10 dBm (orange) and 20 dBm (red). The
   diffusion length from the exponential fits is close to 2 nm. The
   significant difference between the signal extrapolated to zero NiO
   thickness and that of the pure YIG$|$Pt sample, gives an estimate
   interface YIG$|$NiO transparency of 0.05.}
 \label{fig3}
\end{figure}

\begin{figure}
\includegraphics[scale=0.5, angle=0]{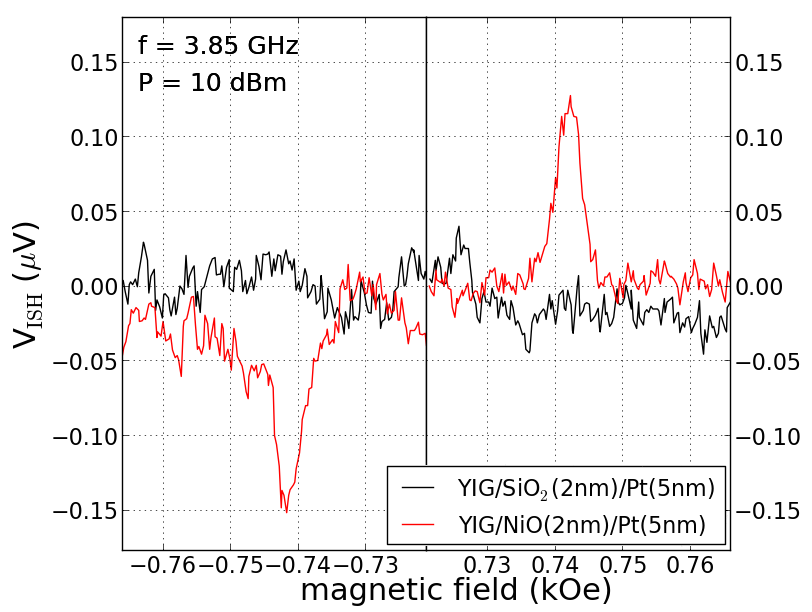}
\caption{Comparison of the effect of the insertion of a 2 nm layer of
  NiO and SiO$_2$ between source and sensing layer demonstrating that
  SiO$_2$ is an efficient angular momentum insulator.}
\label{fig4}
\end{figure}

\textbf{Acknowledgments.}  This research was supported by the French
ANR Grant Trinidad (ASTRID 2012 program) as well as the 'Triangle de
la Physique'.

\textbf{Author contribution.} MV, GdL, OK, CH and JBY planned and
designed the experiment. JBY produced and characterized the
samples. CH, VN and GdL performed the experiments. MV, GdL and OK
wrote the manuscript. All authors discussed the results and commented
on the manuscript.

\textbf{Correspondence.} Correspondence and requests for materials
should be addressed to MV (e-mail: michel.viret@cea.fr).



\end{document}